\documentclass[twoside]{ilcws08}
\usepackage[latin1]{inputenc}
\usepackage[dvips]{graphicx,epsfig,color}
\usepackage{wrapfig,rotating}
\usepackage{amssymb,amsmath,array}
\usepackage{url}
\begin{document}
\title{Measuring Higgs boson associated Lepton Flavour Violation 
in electron-photon collisions at the ILC~\footnote{This proceeding paper is
based on Ref~\cite{Ref:KT}.}} 
\author{Shinya Kanemura$^1$,
    and
        Koji Tsumura$^2$
\vspace{.3cm}\\
1- Department of Physics, University of Toyama,
3190 Gofuku, Toyama 930-8555, Japan \\
2- The Abdus Salam ICTP of
UNESCO and IAEA, Strada Costiera 11, 34151 Trieste, Italy \\
}

\maketitle

\begin{abstract}
%
We study the LFV Higgs production processes
$e^-\gamma\to\ell^-\varphi\;(\ell=\mu,\tau; \varphi=H, A)$
as a probe of Higgs mediated LFV couplings at an electron-photon collider,
where $H$ and $A$ are extra CP even and odd Higgs bosons, respectively,
in the two Higgs doublet model.
Under the constraints from the current data of muon and tau rare decay,
the cross section can be significantly large.
It would improve the experimental upper bounds
on the effective LFV coupling constants.
In addition, the chirality nature of the LFV Higgs coupling constants
can be measured by selecting electron beam polarizations.
\end{abstract}

\section{Introduction}
Lepton Flavour Violation (LFV) is clear evidence of new physics
beyond the standard model (SM). It can be naturally induced in various
new physics scenarios such as supersymmetric extensions of the SM.
The origin of LFV would be related to the structure of the fundamental theory
at high energies. Therefore, new physics models can be explored by
measuring the LFV processes.
In the minimal supersymmetric SM with heavy right-handed neutrinos
(MSSMRN), the LFV Yukawa interactions can be radiatively generated via
the slepton mixing~\cite{Ref:YnuRGE,Ref:slmix}.
The slepton mixing can be induced by the running effect
from the neutrino Yukawa interaction even when flavour blind structure
is realized at the grand unification scale~\cite{Ref:YnuRGE}.

The experimental bound on the effective LFV Yukawa couplings
have been studied extensively~\cite{Ref:DT,Ref:BKDERB,Ref:Sher,Ref:DNR,Ref:KOT}.
These constraints will be improved at PSI MEG~\cite{Ref:MEG},
J-PARC COMET~\cite{Ref:COMET} and Fermilab Mu2e~\cite{Ref:Mu2e} experiments via muon rare decays,
and at CERN LHCb~\cite{Ref:LHCb} and KEK super-B factory~\cite{Ref:SuperB}
via tau rare decays.
In addition, collider signatures of the LFV phenomena
have also been investigated at the CERN Large Hadron Collider
(LHC)~\cite{Ref:LHC}, the International Linear Collider
(ILC)~\cite{Ref:ILC}, and the Neutrino Factory~\cite{Ref:nuFact}.
These collider experiments would be useful to test the
Higgs-boson-associated LFV couplings~\cite{Ref:DT,Ref:DNR,Ref:KOT,Ref:LFVhdecayLHC,Ref:KMOSTT,Ref:KKKO}.

In this report, we discuss the physics potential of the LFV Higgs
boson production process
$e^- \gamma\to \ell^- \varphi\; (\ell=\mu, \tau; \varphi=h, H, A)$
where $h$, $H$ and $A$ are neutral Higgs bosons.
It can be an useful tool for measuring Higgs-boson-mediated LFV parameters
in two Higgs doublet models (THDMs) including
Minimal Supersymmetric SMs (MSSMs).
The total cross sections for these processes can be large for
allowed values of the LFV couplings under the constraint
from the current experimental data.
Measuring these processes, the bounds for the Higgs boson associated
LFV coupling constants can be improved significantly. Furthermore,
the chirality of these couplings can be measured by using the polarized
initial electron beam.

\section{Higgs boson associated LFV coupling constants}
The effective Yukawa interaction for charged leptons
is given in the general framework of the THDM by~\cite{Ref:KOT}
\begin{align}
\mathcal{L}_{\text{lepton}}
=& -\overline{{\ell_R}_i}\left\{Y_{\ell_i}\delta_{ij}\Phi_1
+\left(Y_{\ell_i}\epsilon_{ij}^L+\epsilon_{ij}^RY_{\ell_j}\right)
\Phi_2\right\}\cdot L_j
+\text{H.c.},
\label{Eq:yukawa}
\end{align}
where ${\ell_R}_i (i=1$--$3)$ represent isospin singlet fields of
right-handed charged leptons, $L_i$ are isospin doublets of
left-handed leptons, $Y_{\ell_i}$ are the Yukawa coupling constants
of $\ell_i$, and $\Phi_1$ and $\Phi_2$ are the scalar iso-doublets
with hypercharge $Y=1/2$.
Parameters $\epsilon_{ij}^X (X=L,R)$ can induce LFV interactions
in the charged lepton sector in the basis of the mass eigenstates.
In Model II THDM~\cite{Ref:HHG}, $\epsilon_{ij}^X$ vanishes at the
tree level, but it can be generated radiatively by new physics
effects~\cite{Ref:slmix}.
The effective Lagrangian can be rewritten in terms of physical Higgs
boson fields.
Assuming the CP invariant Higgs
sector, there are two CP even Higgs bosons $h$ and $H$
$(m_h^{}<m_H^{})$, one CP odd state $A$ and a pair of charged
Higgs bosons $H^\pm$.
From Eq.~\eqref{Eq:yukawa}, interaction terms can be deduced to~\cite{Ref:slmix,Ref:KOT}
\begin{align}
&\mathcal{L}_{e\text{LFV}}
= -\frac{m_{\ell_i}}{v\cos^2\beta}
\left(\kappa^L_{i1}\overline{\ell_i}\text{P}_L^{}e
+\kappa^R_{1i}\overline{e}\text{P}_L^{}\ell_i\right)
\left\{\cos(\alpha-\beta)h+\sin(\alpha-\beta)H-i\,A\right\}
+\text{H.c.},
\end{align}
where $\text{P}_L^{}$ is the projection operator to the
left-handed fermions, $m_{\ell_i}$ are mass eigenvalues of charged leptons,
$v=\sqrt{2} \sqrt{\langle\Phi_1^0\rangle^2+\langle\Phi_2^0\rangle^2}$
($\simeq 246$ GeV),
$\alpha$ is the mixing angle between the CP even Higgs bosons, and
$\tan\beta\equiv\langle\Phi_2^0\rangle/\langle\Phi_1^0\rangle$.

Once a new physics model is assumed, $\kappa_{ij}^X$ can be predicted
as a function of the model parameters. In supersymmetric SMs,
LFV Yukawa coupling constants can be radiatively generated
by slepton mixing.
Magnitudes of the LFV parameters
$\kappa^{X}_{ij}$ can be calculated as a function of the parameters
of the slepton sector. For the scale of the dimensionful parameters
in the slepton sector to be of TeV scales, we typically obtain
$|\kappa^{X}_{ij}|^2 \sim (1$--$10) \times 10^{-7}$~\cite{Ref:YnuRGE,Ref:slmix}.
In the MSSMRN only $\kappa^L_{ij}$ are generated by the
quantum effect via the neutrino Yukawa couplings assuming
flavour conservation at the scale of right-handed neutrinos.

Current experimental bounds on the effective LFV parameters
$\kappa_{ij}^X$ are obtained from the data of
non-observation for various LFV processes~\cite{Ref:Rare}.
For $e$--$\tau$ mixing, we obtain the upper bound
from the semi-leptonic decay $\tau\to e\eta$~\cite{Ref:Sher};
$|\kappa^L_{31}|^2+|\kappa^R_{13}|^2\lesssim 6.4\times 10^{-6}
(\frac{50}{\tan\beta})^6(\frac{m_A^{}}{350\text{GeV}})^4,$
for $\tan\beta \gtrsim 20$ and $m_A^{} \simeq m_H^{} \gtrsim 160$ GeV (with
$\sin(\beta-\alpha)\simeq 1$).
The most stringent bound on $e$--$\mu$ mixing is derived from
$\mu\to e\gamma$ data~\cite{Ref:PDG} as
$(4/9)|\kappa^L_{21}|^2+|\kappa^R_{12}|^2\lesssim 4.3\times 10^{-4}
(\frac{50}{\tan\beta})^6
(\frac{m_A^{}}{350\text{GeV}})^4,$
for $\tan\beta \gtrsim 20$ and $m_A^{} \simeq m_H^{} \gtrsim 160$ GeV (with
$\sin(\beta-\alpha)\simeq 1$).
The upper bound on $(4/9)|\kappa_{21}^L|^2+|\kappa_{12}^R|^2$ is expected to
be improved at future experiments such as MEG and COMET for rare muon decays
by a factor of $10^{2\text{--}3}$, while that on
$|\kappa_{31}^L|^2+|\kappa_{13}^R|^2$ is by $10^{1\text{--}2}$
at LHCb and SuperKEKB via rare tau
decays~\cite{Ref:MEG,Ref:COMET,Ref:LHCb,Ref:SuperB}.

\begin{figure*}[tb]
\includegraphics[width=7cm]{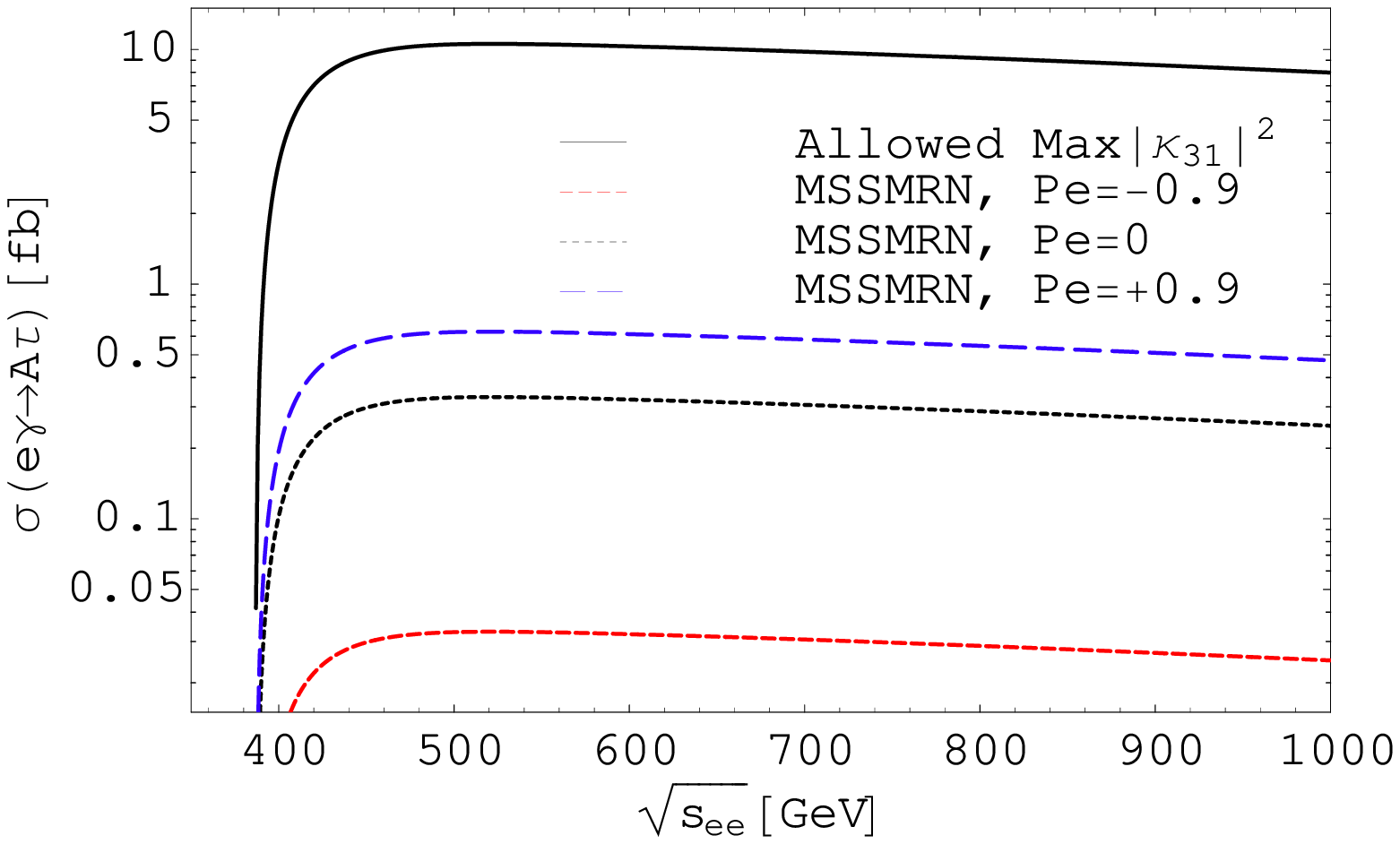}
\includegraphics[width=7cm]{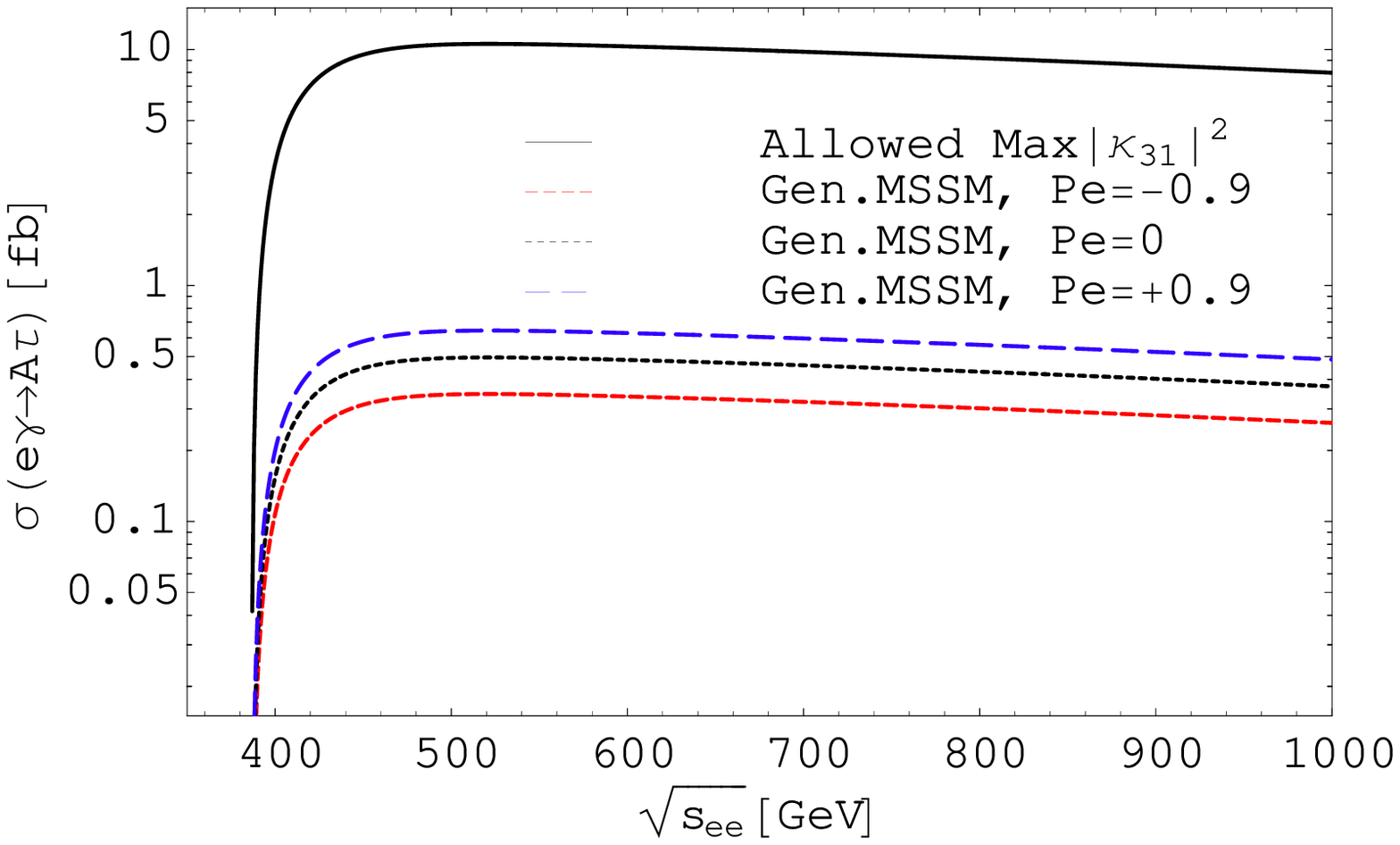}
\caption{
The production cross section of $e^-\gamma \to \tau^- A$ as a function of
the center-of-mass energy  $\sqrt{s_{ee}}$ of the electron-electron system.
Solid curve represents the result in the THDM with the maximal allowed value of
$|\kappa_{31}|^2$ under the current experimental data
in both figures.
 }
\label{FIG:EGamAtau}
\end{figure*}

\section{LFV Higgs production processes}
We now discuss the lepton flavour violating Higgs boson production
processes $e^-\gamma \to \ell^- \varphi\;(\ell=\mu, \tau; \varphi=h, H, A)$
in $e\gamma$ collisions.
The differential cross section is calculated by using the
effective LFV parameters $\kappa_{ij}^X$  as
\begin{align}
%
&\frac{d\widehat{\sigma}_{e^-\gamma\to\ell^-_i\varphi}(\sqrt{s_{e\gamma}})}{d\cos\theta}
=\frac{G_F\alpha_\text{EM}^{}m_\ell^2\beta_{\ell\varphi}}{16\sqrt2s_{e\gamma}}
\frac{\left|\kappa_{i1}\right|^2}{\cos^4\beta}
\frac{\eta_-(\eta_+^2+4z^2)-16z\,m_\ell^2/s_{e\gamma}}{\eta_-^2},
\end{align}
where
$z=(m_{\ell_i}^2-m_\varphi^2)/s_{e\gamma}$ and
$\beta_{\ell\varphi}=\sqrt{\lambda(m_{\ell_i}^2/s_{e\gamma},
m_\varphi^2/s_{e\gamma})}$
with $\lambda(a,b)=1+a^2+b^2-2a-2b-2a b$.
The functions are defined as $\eta_\pm=1+z\pm\beta_{\ell\varphi}\cos\theta$
where $\theta$ is the scattering angle of the outgoing lepton from
the beam direction.
The effective LFV parameters
can be written by
\begin{align}
|\kappa_{i1}|^2 = &\left[  |\kappa^L_{i1}|^2 (1-P_e)
+ |\kappa^R_{1i}|^2 (1+P_e)\right] \times
\begin{cases}
\cos^2(\alpha-\beta)&\text{ for }h\\
\sin^2(\alpha-\beta)&\text{ for }H\\
1&\text{ for }A\end{cases},
\end{align}
where $P_e$ is the polarization of the incident electron beam:
$P_e=-1$ ($+1$) represents that electrons in the beam are
$100\%$ left- (right-) handed.

At the ILC, a high energy photon beam can be obtained by Compton
backward-scattering of laser and an electron beam~\cite{Ref:PLC}.
The full cross section can be evaluated from that for the sub process by convoluting
with the photon structure function as~\cite{Ref:PLC}
\begin{align}
\sigma\left(\sqrt{s_{ee}}\right)
=\int_{x_{min}}^{x_{max}}dx\,F_{\gamma/e}(x)\,
\widehat{\sigma}_{e^-\gamma\to\ell^-\varphi}
(\sqrt{s_{e\gamma}}),
\end{align}
where $x_{max}=\xi/(1+\xi)$, $x_{min}=(m_\ell^2+m_\varphi^2)/s_{ee}$,
$\xi=4E_e\omega_0/m_e^2$ with $\omega_0$ to be the frequency of the
laser and $E_e$ being the energy of incident electrons,
and $x=\omega/E_e$ with $\omega$ to be the photon energy in the scattered photon beam.
The photon distribution function is given in Ref.~\cite{Ref:PLC}.
We note that when $\sin(\beta-\alpha)\simeq 1$ and $m_H^{} \simeq m_A^{}$
(In the MSSM, this automatically realizes for $m_A \gtrsim 160$ GeV)
signal from both $e^-\gamma\to\ell^-H$
and $e^-\gamma\to\ell^-A$ can be used to measure the LFV
parameters, while the cross section for $e^-\gamma\to\ell^-h$ is
suppressed.

In FIG.~\ref{FIG:EGamAtau}, we show the full cross sections of
$e^-\gamma \to \tau^- A$ as a function of the center-of-mass energy
of the $e^-e^-$ system for $\tan\beta=50$ and $m_A^{}=350$ GeV.
Scattered leptons mainly go into the forward direction, however
most of events can be detected by imposing the escape cut
$\epsilon\le\theta\le\pi-\epsilon$
where $\epsilon=20$ mrad~\cite{Ref:Anipko}.
The cross section can be around $10$ fb with the maximal allowed values
for $|\kappa_{31}|^2$ under the constraint from the $\tau\to e\eta$ data.
The results correspond that, assuming the integrated luminosity of the $e\gamma$
collision to be $500$ fb$^{-1}$ and the tagging efficiencies of a $b$
quark and a tau lepton
to be $60\%$ and $30\%$, respectively,
about $10^3$ of $ \tau^- b\bar b$ events can be observed as the signal,
where we multiply factor of two by adding both
$e^-\gamma\to\ell^-A\to\ell^-b\overline{b}$ and
$e^-\gamma\to\ell^-H\to\ell^-b\overline{b}$.
Therefore, we can naively say that non-observation of the signal
improves the upper bound for the $e$-$\tau$ mixing by $2$--$3$ orders of
magnitude if the backgrounds are suppressed.
In FIG.~\ref{FIG:EGamAtau} (left), those with a set of the typical values of
$|\kappa^L_{31}|^2$ and $|\kappa^R_{13}|^2$ in the MSSMRN are shown for
 $P_e=-0.9$ (dashed), $P_e=+0.9$ (long dashed), and $P_e=0$ (dotted), where we take
 $(|\kappa^L_{31}|^2, |\kappa^R_{13}|^2)=(2\times 10^{-7}, 0)$.
The cross sections are sensitive to the polarization of
the electron beam. They can be as large as $0.5$ fb for $P_e=-0.9$,
while it is around $0.03$ fb for $P_e=+0.9$.
In FIG.~\ref{FIG:EGamAtau} (right), the results with
$(|\kappa^L_{31}|^2, |\kappa^R_{13}|^2)=(2\times 10^{-7}, 1\times
 10^{-7})$ in general supersymmetric models
 are shown for each polarization of the incident electrons.
The cross sections are a few times $1$ fb and not sensitive for polarizations.
Therefore, by using the polarized beam of the electrons
we can separately measure $|\kappa_{31}^L|^2$ and $|\kappa_{13}^R|^2$ and
 distinguish fundamental models with LFV.

\begin{figure*}[tb]
\includegraphics[width=7cm]{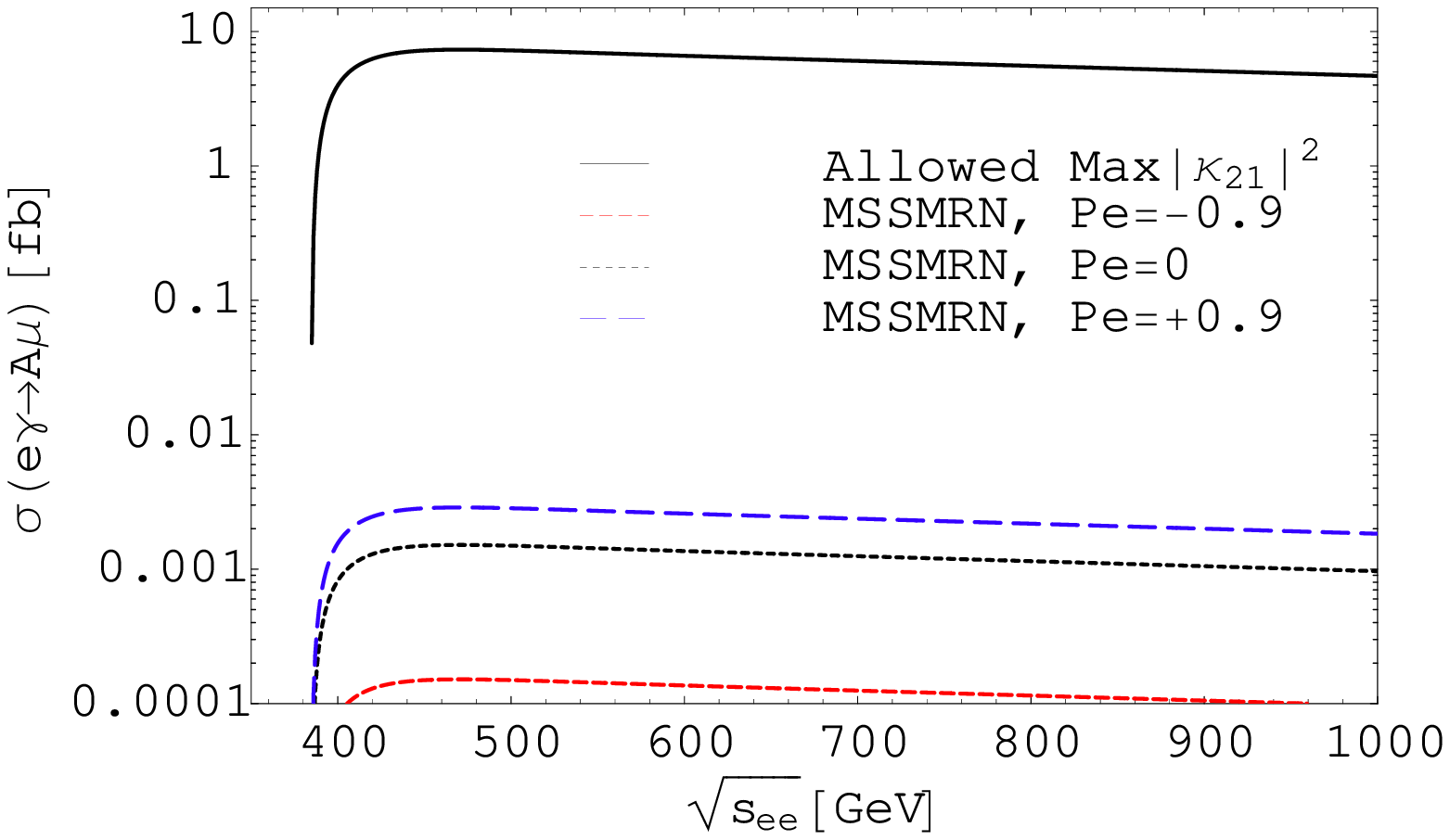}
\includegraphics[width=7cm]{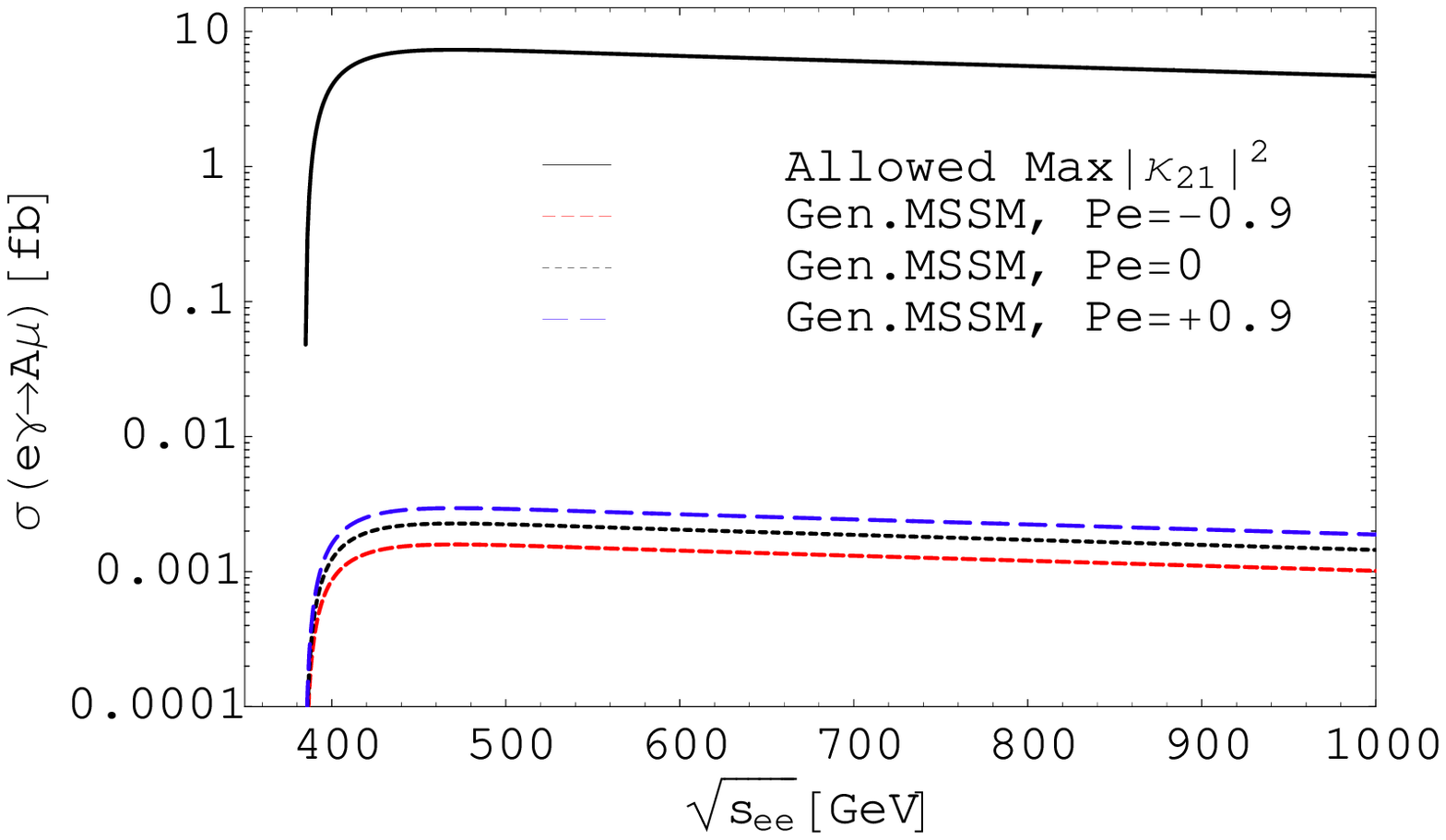}
\caption{
The production cross section of $e^-\gamma \to \mu^- A$ as a function of
the center-of-mass energy  $\sqrt{s_{ee}}$ of the electron-electron system.
Solid curve represents the result in the THDM with the maximal allowed value of
$|\kappa_{21}|^2$ under the current experimental data
in both figures.
}
\label{FIG:EGamAmu}
\end{figure*}
In FIG.~\ref{FIG:EGamAmu}, the full cross sections of $e^-\gamma \to
\mu^- A$ are shown for $\tan\beta=50$ and $m_A^{}=350$ GeV.
Those with the maximally allowed values for
$|\kappa_{21}|^2=|\kappa^L_{21}|^2+|\kappa^R_{12}|^2$
from the $\mu \to  e \gamma$ data
can be $7.3$ fb
where we here adopted the same escape cut as before discussed~
\footnote{If $10$ mrad for the cut is taken instead of $20$ mrad, the
numbers of events are slightly enhanced; $10.6$ fb to $11.0$ fb
($7.3$ fb to $8$ fb) for the $\tau$-$\varphi$  ($\mu$-$\varphi$) process.}.
This means that about a few times $10^3$ of the signal $\mu^- b\bar b$ can be
produced for the integrated luminosity of the $e\gamma$ collision to be
$500$ fb$^{-1}$, assuming tagging efficiencies to be $60\%$ for a $b$
quark and $100\%$ for a muon, and using both $e^-\gamma \to
\mu^- A$ and $e^-\gamma \to \mu^- H$.
These results imply that $e\gamma$ collider can improve the bound on
the $e$-$\mu$ by a factor of $10^{2-3}$.
Obtained sensitivity can be as large as those at
undergoing MEG and projected COMET experiments.
Because of the different dependencies on the parameters in the model,
$\mu\to e\gamma$ can be sensitive than the LFV Higgs boson production
for very high $\tan\beta (\gtrsim 50)$ with fixed Higgs boson mass.
We also note that rare decay processes can measure the effect of other
LFV origin when Higgs bosons are heavy. Therefore, both the direct and
the indirect measurements of LFV processes are complementary to each other.
In FIG.~\ref{FIG:EGamAmu} (left), those in the MSSMRN are shown for
 $P_e=-0.9$ (dashed), $P_e=+0.9$ (long dashed), and $P_e=0$ (dotted), where we take
 $(|\kappa^L_{21}|^2, |\kappa^R_{12}|^2)=(2\times 10^{-7}, 0)$.
They can be as large as a few times $10^{-3}$ fb for $P_e=-0.9$ and
$P_e=0$, while it is around $10^{-4}$ fb for $P_e=+0.9$.
In FIG.~\ref{FIG:EGamAmu} (right), the results with
$(|\kappa^L_{21}|^2, |\kappa^R_{12}|^2)=(2\times 10^{-7}, 1\times
10^{-7})$ are shown in general supersymmetric models in a similar
manner.

It is understood that these processes are clear against backgrounds.
For the processes of  $e^-\gamma \to \tau^- \varphi \to \tau^- b\bar b$.
The tau lepton decays into various hadronic and
leptonic modes.
The main background comes from $e^-\gamma\to W^-Z\nu$, whose cross
section is of the order of $10^2$ fb.
The backgrounds can strongly be suppressed
by the invariant mass cut for $b\bar b$.
The backgrounds for the
process $e^-\gamma \to \mu^- \varphi \to \mu^- b\bar b$ also comes from
$e^-\gamma\to W^-Z\nu\to \mu^-b\overline{b}\nu\overline{\nu}$ which
is small enough. Signal to background ratios are better than
${\mathcal O}(1)$ before kinematic cuts. They are easily
improved by the invariant mass cut, so that our signals can be
almost background free.

\section{Conclusion}
We have studied the Higgs boson associated LFV at an electron photon collider.
Lots of new physics model can predict the LFV Yukawa interactions.
The cross section for 
$e^-\gamma\to\ell^-\varphi\;(\ell=\mu,\tau; \varphi=H, A)$
can be significant for the allowed values of the effective LFV couplings
under the current experimental data.
By measuring these processes at the ILC,
the current upper bounds on the effective LFV Yukawa coupling constants
are expected to be improved in a considerable extent.
Such an improvement can be better than those at MEG and COMET experiments
for the $e$-$\mu$-$\varphi$  vertices, and those at LHCb and SuperKEKB
for the $e$-$\tau$-$\varphi$ vertices.
Moreover, the chirality of the LFV Higgs coupling can be separately measured
via these processes by using the polarized electron beam.
The electron photon collider can be an useful tool of measuring Higgs boson
associated LFV couplings.

\vspace{5mm}
\noindent {\bf Acknowledgments}\\
The authors would like to thank the members of the ILC physics
subgroup~\cite{Ref:WG} for useful discussions.
The work of S.K. was supported, in part, by Grant-in-Aid, Ministry of Education,
Culture, Sports, Science and Technology, Government of Japan, No.~18034004.


\begin{footnotesize}



%

\end{footnotesize}


\end{document}